\documentclass[12pt]{iopart}

\usepackage{graphicx}
\usepackage{xcolor}
\begin{document}

\title[]{Spin-dependent scattering and magnetic proximity effect in {Ni}-doped {Co}/{Cu} multilayers as a probe of atomic magnetism}

\author{Yu~O~Tykhonenko-Polishchuk$^{1,2}$, D~M~Polishchuk$^{1,2}$, T~I~Polek$^2$, D~D~Yaremkevych$^2$, A~F~Kravets$^{1,2}$, A~I~Tovstolytkin$^2$, A~N~Timoshevskii$^3$, V~Korenivski$^1$}

\address{$^1$Nanostructure Physics, Royal Institute of Technology, 10691 Stockholm, Sweden}
\address{$^2$Institute of Magnetism, National Academy of Sciences of Ukraine and Ministry of Education and Science of Ukraine, 36~b Vernadsky Blvd., 03142 Kyiv,
Ukraine}
\address{$^3$Institute for Metal Physics, National Academy of Sciences of Ukraine, 36 Vernadsky Blvd., 03680 Kyiv, Ukraine}
\ead{yuliatp@kth.se}

\vspace{10pt}
\begin{indented}
\item[]July 2018
\end{indented}

\begin{abstract}
We investigate the spin transport and ferromagnetic resonance properties of giant magnetoresistive (GMR) Co/Cu-Ni multilayers with variable levels of Ni doping in the Cu spacer. We present an experimental evidence for a magnetic-to-diamagnetic transition in the atomic magnetic moment of Ni in the Cu matrix for concentrations below 15~at.~\%~Ni. As its concentration is increased, Ni atoms turn into spin scattering centers, which is manifested experimentally as a step-like change in the GMR of the multilayers. This behavior is observed in multilayers with gradient-doped Cu spacers, where only the inner region was doped with Ni. In the uniformly doped spacers the GMR decreases monotonously with increasing Ni content, indicating that Ni atoms are magnetic and act as spin relaxation centers in the entire dopant-concentration range studied. We explain the difference in the observed GMR behavior as due to a strong magnetic proximity effect in the uniform spacers, which is efficiently suppressed in the gradient spacers. The observed magnetic phase transition is fully supported by our detailed ab-initio calculations, taking into consideration structural relaxation in the system as well as potential Ni clustering. Controlling the loss or gain of the atomic magnetism for a specific dopant can be a tool in probing and controlling spin relaxation in materials and devices for spin-valve and spin-torque based applications.
\end{abstract}

\submitto{\JPCM}

\vspace{2pc}
\noindent{\it Keywords}: spin-transport properties, magnetic multilayers, giant magnetoresistance, magnetic proximity effect, diamagnetism.

\maketitle
%
%

\section{Introduction}

Magnetotransport measurements is an effective tool in characterizing the intrinsic magnetic properties of nanostructured materials. A classic example is the giant magnetoresistance (GMR) in magnetic multilayers \cite{Baibich1988, Binasch1989}, often the most sensitive probe of spin-dependent electron scattering in a given system, such for example as a multilayer exhibiting the oscillatory Ruderman-Kittel-Kosuya-Yosida (RKKY) interaction \cite{Parkin1990}. Other recent prominent examples include the anisotropic magnetoresistance used to study antiferromagnet-based systems \cite{Marti2014} and a variety of spin-Hall effects \cite{Sinova2015, Zelezny2018} at nonmagnetic/(anti-)ferromagnetic metal \cite{Miron2011, Liu2012} or insulator \cite{Huang2012, Nakayama2013} interfaces. Here we use spin transport to probe subtle changes in impurity magnetism that cannot be observed by conventional magnetometry. We exploit the fact that a magnetic impurity in the spacer of a magnetic multilayer is a spin-scattering center and can significantly reduce the GMR effect for impurity concentrations as low as a few percent \cite{Yang1994}. In contrast, a nonmagnetic (diamagnetic, with no spontaneous magnetic moment) impurity contributes only to momentum scattering, which is spin-independent and does not affect the GMR \cite{Yang1994, Bass1995}.

There are several magnetic binary alloys, among which Cu$_{100-x}$Ni$_x$ and V$_{100-x}$Fe$_x$ are the most known, with diamagnetic states in their concentration diagrams for the nominally magnetic impurities of Ni and Fe, respectively \cite{Bozorth1951}. As $x$ is increased, the impurity should become paramagnetic and eventually ferromagnetic as $x$ approaches~1. In particular, Cu$_{100-x}$Ni$_x$ alloys are a text-book system, where Ni has been predicted to gradually lose its atomic magnetic moment (ionic magnetic moment, more strictly speaking) and become diamagnetic at low concentrations. Though being extensively studied theoretically \cite{Robbins1969, Teeriniemi2015}, the experimental investigation of the magnetic state of Ni ions in a Cu matrix using conventional magnetometry or neutron scattering techniques \cite{Ahern1958, Hicks1969, Straumal2011} faces great difficulties in terms of sensitivity to  low-volume atomic-scale magnetic transformations (quasi-2D in thin films). In this work, by layering nanometer-thin Cu$_{100-x}$Ni$_x$ alloys with strongly ferromagnetic Co, we exploit the high sensitivity of the GMR to the spin state of the Ni-impurities and vary the Ni-concentration to explore the magnetic phase diagram of the system.

The effect of alloying the spacer with magnetic impurities on the GMR properties was studied earlier \cite{Yang1994, Bass1995, Parkin1993, Okuno1993, Bobo1993}. In particular, Co multilayers with Ni-doped Cu spacers revealed significant changes in the position and oscillation period of the antiferromagnetic peaks of the RKKY interlayer coupling  \cite{Parkin1993, Okuno1993, Bobo1993}. This was explained in terms of an impurity-modified Fermi-surface topology. These studies, however, left without attention the strong \emph{magnetic proximity effect} at the ferromagnet/paramagnet interfaces \cite{Samant1994, Hernando1995, Navarro1996, Kravets2012} -- direct proximity of the paramagnetic atoms in the uniformly doped spacer to the ferromagnetic interfaces results in strong exchange-induced magnetization, decaying off the interfaces into the spacer. It is important to point out that such exchange bias of paramagnetic impurities reduces the effective thickness of the spacers as well as the spin-scattering properties of the interfaces, thereby significantly altering the observed GMR. In order to exclude the unwanted proximity-related effects, we found it critical to use the so-called gradient doping \cite{Kravets2012}, where the direct exchange interaction between the ferromagnet and the dilute magnetic spacer is broken by incorporating a thin nonmagnetic interface of pure~Cu.

In this work, we show experimentally, by studying the GMR of Co/Cu multilayers with gradient Ni-doped Cu-spacers as a function of Ni content, that a magnetic-to-diamagnetic transition takes place for low concentrations of atomic Ni in Cu. Great care is taken to eliminate the magnetic proximity effect, which is found to fully suppress the magnetic phase transition in samples with uniformly doped spacers. 

\section{Experimental details}

\begin{figure}[t]
\centering
\includegraphics[width=8cm]{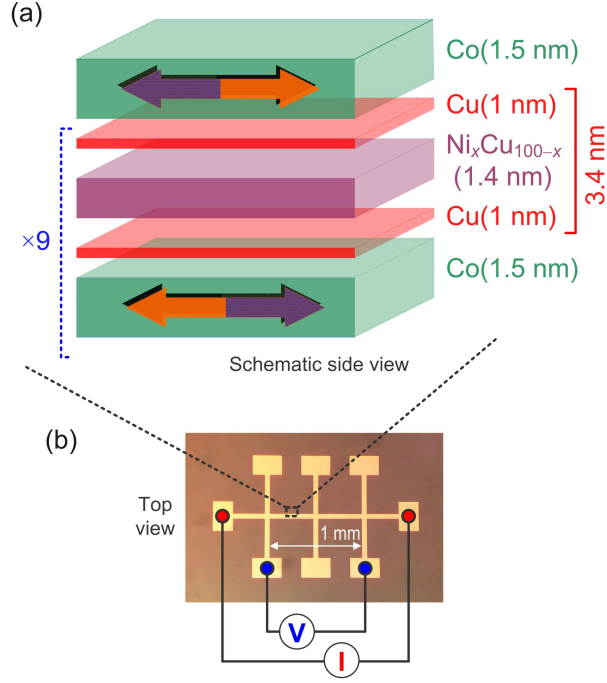}
\caption{(a) Layout of multilayers with gradient-doped spacers, Cu/Cu$_{100-x}$Ni$_x$/Cu. Samples with uniformly-doped Cu$_{100-x}$Ni$_x$ spacers (not shown) have total spacer thickness equal to that of gradient spacer (3.4 nm). Arrows represent antiparallel magnetization orientation of Co layers corresponding to maximum resistance of stack. (b) Optical image of photo-lithographically patterned multilayer films, with schematic of four-probe configuration for magneto-transport measurements.}
\label{fig_1}
\end{figure}

Two series of multilayers Co(1.5)/[Cu$_{100-x}$Ni$_x$(3.4)/ Co(1.5)]$_{\times{9}}$ (with uniform spacers) and Co(1.5)/[Cu(1)/Cu$_{100-x}$Ni$_x$(1.4)/Cu(1)/Co(1.5)]$_{\times{9}}$ (with gradient spacers), as shown in \fref{fig_1}(a), with $x$ = 0-70~at.~\%~Ni, were grown on oxidized Si substrates at room temperature using a dc magnetron sputtering system (AJA Inc.). Layers of Cu$_{100-x}$Ni$_x$ binary alloys of varied composition were deposited using co-sputtering from separate Ni and Cu targets. The alloy composition was varied by setting the deposition rates of the individual Ni and Cu components based on calibrations obtained by thickness profilometry. The magnetic properties were characterized using a vibrating-sample magnetometer (VSM, by Lakeshore Cryogenics). The ferromagnetic resonance (FMR) measurements were carried out using an X-band ELEXSYS E500 spectrometer (Bruker) at a constant operating frequency of 9.88 GHz and in-plane applied magnetic field. The magneto-resistance measurements were performed in the current-in-plane configuration using the four-probe technique on photo-lithographically patterned multilayer samples with 1$\times$0.05 mm planar dimensions, as shown in \fref{fig_1}(b).

First-principles calculations of the Ni-atomic magnetism in Cu-Ni alloys were performed within the density functional theory (DFT) using the FLAPW method (Wien2k package \cite{Blaha2001}). The calculations were performed taking into account spin polarization in the collinear approximation, and the GGA exchange correlation potential was taken to be of the form described in \cite{Perdew1996}. In one simulation series, structural relaxation was performed for the lattice parameters (fcc) and the atomic positions, with the Ni atoms distributed at random within the alloy. The second simulation series reflected the gradient spacer, where the Cu-Ni alloy was grown on a pure Cu layer, with a fixed lattice constant for the alloy equal to that of fcc Cu as well as random Ni atomic configuration. The third simulation series used the fcc-Cu lattice constant for the Cu-Ni alloy, but took into account clustering of Ni atoms. For all simulation, the local atomic magnetic moment for a given Ni site was found to be a function of the Ni-concentration and, more specifically, the number of the nearest-neighbour Ni atoms. The exchange integrals were obtained for two coordination spheres of the fcc lattice. More details on the theoretical method can be found in \cite{Kravets2012, Kravets2012a}.

\section{Results and Discussion}

\begin{figure}
\centering
\includegraphics[width=8.5cm]{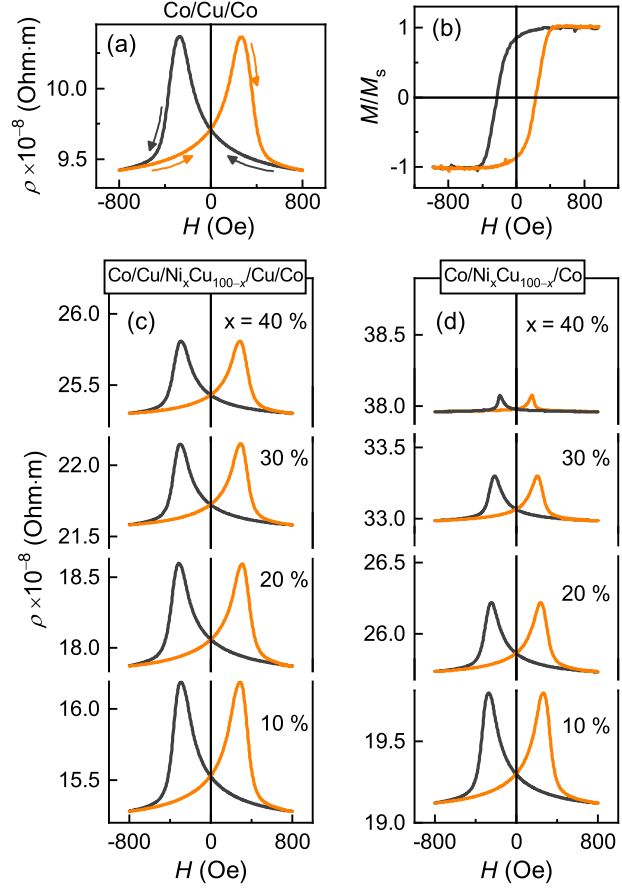}
\caption{Field dependence of resistivity for Co/Cu multilayer with no Ni impurities in Cu spacer (a) and corresponding magnetization loop (b). Magneto-resistivity for gradient (c) and uniform (d) Ni-doped spacer series, for various Ni concentrations ($x$).}
\label{fig_2}
\end{figure}

\Fref{fig_2}(a) shows the measured resistance versus magnetic field data for a reference (un-doped, pure-Cu spacer) multilayer, which exhibits the typical GMR shape, with two resistivity peaks at the coercive fields of the respective magnetization loop shown in \fref{fig_2}(b). The peak height decreases with increasing the concentration of Ni in the Cu spacer, as seen in the data of \fref{fig_2}(c),(d) for the gradient and uniformly doped spacer samples, respectively. The decrease in the GMR is much more pronounced in the samples with uniformly doped spacers [\fref{fig_2}(d)]. Also important to notice is that the overall resistivity of the samples increases with increasing concentration of the Ni impurities and this rise is much steeper in the case of the uniform spacer.

\begin{figure}
\centering
\includegraphics[width=8cm]{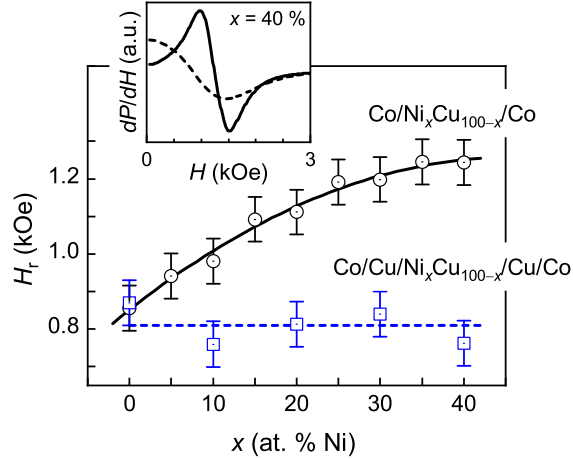}
\caption{In-plane FMR resonance field ($H_\mathrm{r}$) versus Ni-concentration ($x$) for multilayers with gradient (squares) and uniform (circles) spacers. Solid and dashed lines for guiding the eye. Inset shows FMR spectra for uniform- (solid) and gradient-spacer (dashed) samples with $x$ = 40~\%.}
\label{fig_3}
\end{figure}

All FMR spectra contain one resonance line, shown in the inset to \fref{fig_3}, which is ascribed to the response of the ferromagnetic Co layers. \Fref{fig_3} shows the concentration dependences of the in-plane resonance field, $H_\mathrm{r}$, for the two series of samples. $H_\mathrm{r}$ monotonously increases with rising Ni concentration ($x$) for the uniform-spacer series. It is constant, however, for the gradient-spacer series. The observed difference in the behavior of $H_\mathrm{r}$ versus $x$ indicates a change in the magnetic properties of the Co layers having the interfaces exchange-interacting with the Ni impurities, and the absence of such change for gradient spacers where the interfaces are pure Co/Cu.

FMR spectra were measured as a function of in-plane angle and revealed no in-plane magnetic anisotropy, which justifies the use of Kittel's formalism for isotropic ferromagnetic films: ($\omega/\gamma)^2 =  H_\mathrm{r}(H_\mathrm{r} + 4{\pi M_\mathrm{s}}$) \cite{Kittel1949}, where $H_\mathrm{r}$ is the FMR resonance field at a given fixed measurement frequency $f = \omega/2\pi$, $\gamma$ is the gyromagnetic ratio, and $M_\mathrm{s}$ is the saturation magnetization. Within this formalism, the position of the resonance line ($H_\mathrm{r}$) is inversely proportional to the saturation magnetization of the film: an increase in $H_\mathrm{r}$ corresponds to a decrease in $M_\mathrm{s}$ and vice-versa. With rising Ni concentration $x$ in the uniform spacers, the increasing $H_\mathrm{r}$ indicates an effective decrease in $M_\mathrm{s}$ of the magnetic layers. This can be explained in terms of the magnetic proximity effect \cite{Hernando1995, Navarro1996, Kravets2012}, in which the Co interfaces substantially enhance the magnetic polarization of the adjacent Ni impurities, effectively increasing the now Ni/Co/Ni magnetic layers’ thickness. In turn, these exchange-coupled interfacial Ni atoms contribute into the total $M_\mathrm{s}$ of the Co/Ni layers, effectively lowering the total $M_\mathrm{s}$ due to their smaller atomic moment and higher thermal agitation. Thus, uniform Ni doping of the Cu spacer modifies the properties of the Co layers, especially the interfaces, which should be taken into account in the analysis to follow. As expected, in line with the original design, the FMR data for the gradient sample series confirm a complete suppression of the magnetic proximity effect ($H_\mathrm{r}$ independent of $x$), which allows us to study the evolution of the intrinsic atomic magnetism of Ni in Cu.

\begin{figure*}
\centering
\includegraphics[width=14cm]{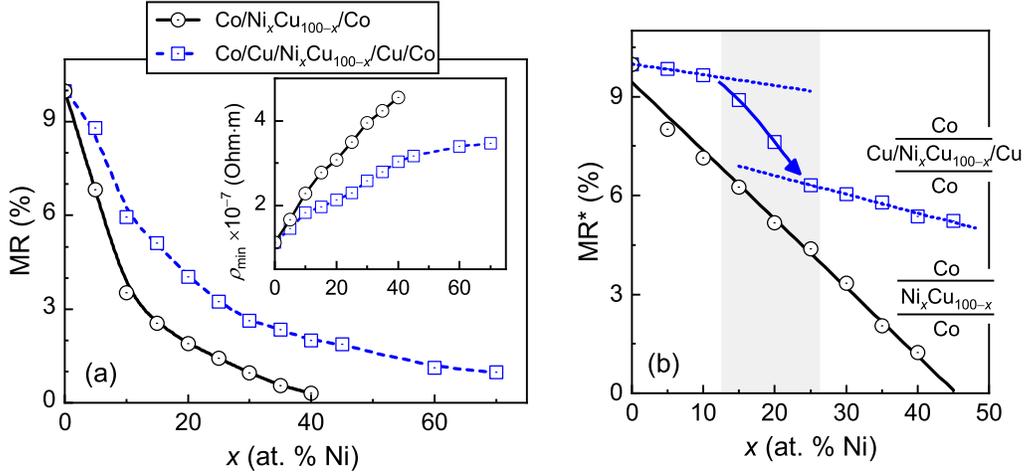}
\caption{(a) Magnetoresistance (MR) extracted from data in \fref{fig_2} using \eref{1}, as a function of Ni concentration for gradient (squares) and uniform (circles) spacers. Inset shows concentration dependence of resistivity at saturation, $\rho_\mathrm{min}$, for both sample series. (b) Magnetoresistance MR$^\ast$ vs $x$ with spin-independent impurity scattering subtracted using \eref{2}. Solid and dashed lines are linear fits to uniform and gradient data sets, respectively.}
\label{fig_4}
\end{figure*}

The concentration dependences of the magnetoresistance (MR) for the two sample series are compared in \fref{fig_4}(a). The MR values were calculated using the GMR convention: 

\begin{equation}
\mathrm{MR} = (\rho_\mathrm{max}/\rho_\mathrm{min} - 1),
\label{1}
\end{equation}

\noindent where $\rho_\mathrm{min}$ and $\rho_\mathrm{max}$ are the minimum and maximum resistivity of $\rho (H)$ shown in \fref{fig_2}(a), which respectively correspond to the saturation and coercive fields of the relevant $M (H)$ loops, such as the one shown in \fref{fig_2}(b). MR vs $x$ in both cases shows a steep decrease in magnitude with increasing Ni-concentration. The MR for the uniform-spacer system vanishes to zero already at about 40~at.~\%~Ni, unlike that for the gradient-spacer multilayers, which remains finite ($\sim1$~\%) for $x$ up to 80~at.~\%~Ni, where the bulk Cu-Ni alloy is ferromagnetic at room temperature \cite{Bozorth1951, Ahern1958}. The vanishing MR and the enhanced magnetization seen in the FMR data point to ferromagnetic ordering in the uniform spacers above 40~at.~\%~Ni. Then, the Co layers exchange-couple through the spacers and the entire multilayer behaves as a single-layer ferromagnetic film, even though the spacer should nominally be paramagnetic for $x$ up to 70~at.~\%~Ni (at room temperature). Such direct-exchange interlayer coupling has been analyzed in detail for the Cu-Ni system \cite{Kravets2012, Kravets2014} and is due to a strong magnetic proximity effect. In contrast, the non-zero MR in the whole range of $x$ = 0-80~at.~\%~Ni, combined with the constant magnetization seen in the FMR data, indicate the presence of well-defined ferromagnetic/nonmagnetic interfaces in the gradient spacers, free from the magnetic proximity effect for the chosen Co/Cu/Cu-Ni gradient interface geometry. The inner Ni-doped spacer is then essentially unperturbed by the Co-proximity and can manifest its intrinsic magnetism via the mechanism of spin-dependent scattering reflected in the measured GMR \cite{Valet1993}. 

The spin-independent part of the multilayer resistivity taken as the resistivity at magnetic saturation, $\rho_\mathrm{min}$, increases with rising Ni-concentration, as shown in the inset to \fref{fig_4}(a). This increase in $\rho_\mathrm{min}$ is attributed to the increased amount of impurity scattering, foremost spin-independent, within the spacers, supported also by the measured difference in $\rho_\mathrm{min}$($x$) for the two sample series: the number of Ni impurities for a given $x$ is larger in the uniform spacers, in which the entire 3.4-nm is the Cu-Ni alloy, whereas only 1.4-nm of the gradient spacer is subject to Ni-doping. The increase in the momentum scattering and thereby resistivity on alloying of pure elements, such as Cu, is well known \cite{Yang1994,Bass1995} and should be subtracted from the total resistivity change in order to single out the spin-dependent contribution. The simplest form for the total resistivity, combining the intrinsic, impurity-spin-independent (momentum), and impurity-spin-dependent contributions is $\rho (x, H) = \rho_0 + \rho_\mathrm{MR}(H) + \Delta\rho_0(x)$. Here $\rho_\mathrm{MR}$ is the spin-dependent resistivity strongly influenced by the applied magnetic field, $\Delta\rho_0(x)$ -- the increase in the saturation resistivity ($\rho_\mathrm{min}$) for a given concentration $x$. Naturally, $\rho_\mathrm{MR}(H \geq H_\mathrm{s})$ = 0, where $H_\mathrm{s}$ is the saturation field, and $\Delta\rho_0(x = 0)$ = 0, which means that $\rho(x = 0,  H \geq H_\mathrm{s}) = \rho_0 \equiv \rho_\mathrm{min}(0)$. Subtracting the spin-independent contribution, $\Delta\rho_0(x)$, from the resistivity, expression \eref{1} can be rewritten as

\begin{equation}
\mathrm{MR}^\ast = [\rho_\mathrm{max}(x) - \rho_\mathrm{min}(x)]/\rho_\mathrm{min}(0),
\label{2}
\end{equation}

\noindent where $\rho_\mathrm{min}$ and $\rho_\mathrm{max}$ are the minimum and maximum resistivity in $\rho(H)$ for a given $x$. 

The Ni-concentration dependence of MR$^\ast$, extracted from the data of \fref{fig_4}(a) using \eref{2} for the uniform-spacer samples, gradually decreases to zero at about 45~at.~\%~Ni, as shown in \fref{fig_4}(b). In contrast, MR$^\ast$($x$) for the gradient-spacer series is not monotonic and shows a clear step at 15-25~at.~\%~Ni. This concentration interval is the region where Ni in Cu has been predicted to transition from essentially a nonmagnetic state to having a strong atomic magnetic moment \cite{Robbins1969, Kravets2012}, which would make it a strong spin scattering center. At $x>$ 30~at.~\%~Ni, MR$^\ast$ decreases at a significantly higher pace compared to the low-$x$ region, proportionally to the increase in the number of Ni impurities (spin scatterers).

\begin{figure*}
\centering
\includegraphics[width=8.5cm]{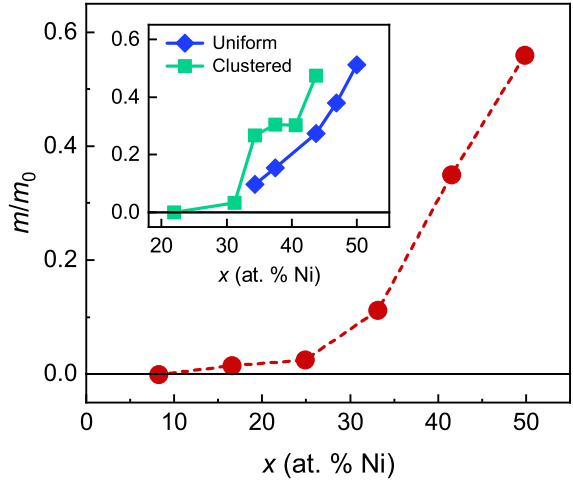}
\caption{Locally averaged atomic magnetic moment of Ni (red) in Cu matrix as a function of Ni-concentration, obtained from ab-initio calculations, taking into account structural relaxation and with randomly distributed Ni atoms. Magnetic moment of Ni is normalized to bulk Ni atomic magnetic moment, $m_0$. Inset shows same as main panel, calculated with fixed lattice constant (pure fcc-Cu): with randomly distributed Ni atoms (blue) and clustered Ni atoms (green).}
\label{fig_5}
\end{figure*}

The results of our ab-initio calculations of the magneto-structural properties of a set of atomic supercells Cu$_{32-\mathrm{n}}$Ni$_\mathrm{n}$ are presented in \fref{fig_5}. We previously have used these structures for modelling the magnetic properties of the Cu$_{100-x}$Ni$_x$ alloy. The main panel of \fref{fig_5} shows the Ni-concentration dependence of the Ni atomic magnetic moment in this alloy obtained as detailed in \cite{Kravets2012}. This dependence was calculated by relaxing the lattice of the model structure for each concentration of Ni and with randomly distributed Ni in Cu. It shows a monotonous decrease in the Ni atomic moment with decreasing $x$, with the atomic moment vanishing to zero in the concentration range of 15 to 25~at.~\%. This fully confirms the interpretation of the experimental data of \fref{fig_4}, pointing to atomically nonmagnetic Ni in Cu below about 15~at.~\%. 

Specifically for the gradient spacer, Cu/Cu$_{100-x}$Ni$_x$/Cu, we note that the formation of the inner Cu-Ni alloy on pure Cu takes place without an interface as such. This means that the lattice parameter of the alloy should be very close to that of Cu for all concentrations in the region of interest (below 50~at.~\%~Ni) and that the alloy is therefore stretched. It is well known that on increasing the Ni content in the bulk Cu-Ni alloy, the equilibrium lattice constant decreases. The Ni-Ni spacing is somewhat larger than in pure Ni, which results in a tendency toward Ni-clustering caused by surface diffusion. We have studied in detail the effect of clustering in dilute ferromagnetic Cu$_{100-x}$Ni$_x$ films obtained by magnetron sputtering in the range of high Ni-concentrations \cite{Kravets2012a}. Our analysis had shown that the observed phase separation is due to the Ni-Ni exchange interactions, which favors clustering of Ni in Cu. The same effect of Ni-clustering should also be expected at $x<$ 50~at.~\%, even though the Ni atomic density decreases in the nearest coordinate spheres. A cluster forming in the concentration range 30~at.~\% $<x<$ 40~at.~\%, has the central Ni atom surrounded by 9 to 12 (maximum for the fcc lattice) Ni neighbors within the first coordinate sphere.

In order to quantify the above arguments, we have performed detailed \textit{ab-initio} modeling of the magneto-structural properties of the Cu$_{100-x}$Ni$_x$ alloy in the concentration interval 20~at.~\% $<x<$ 50~at.~\%. The modeling was carried out using two sets containing 5 and 6 structural configurations of Cu$_{32-\mathrm{n}}$Ni$_\mathrm{n}$. The lattice parameter for all structures was taken as $a = 2a_\mathrm{0}$, where $a_\mathrm{0}$ is the parameter of the fcc lattice of pure Cu. The first set had 5 structures with the Ni atoms distributed randomly (uniformly). The second set had Ni clusters containing 7 and 10-14 Ni atoms. All calculations used optimization of the atomic positions within the unit cell. The results are given in the inset to \fref{fig_5} and show that in this concentration range the Cu$_{100-x}$Ni$_x$ alloy in the gradient spacer Cu/Cu$_{100-x}$Ni$_x$/Cu can have two magneto-structural states. The full energy of these two phases are practically equal (within the numerical uncertainty). The modelling clearly shows a rather steep transition of Ni clusters from a ferromagnetic into a diamagnetic state (green) compared to that for the uniform case (blue). A more detailed analysis would require modeling in the non-collinear spin-spin approximation. We note that the extremely small size of the Ni clusters should make them fully superparamagnetic and therefore efficient spin-flip centers at room temperature (used on the experiment). 

Our modeling is conclusive in that all simulated configurations, with and without structural relaxation, with and without Ni-clustering, point to the same main effect -- a magnetic to nonmagnetic phase transition for Ni in Cu for concentrations below about 15~at.~\%~Ni. The effect is a rather unique magnetic phase transition that has long been predicted, but required a highly-specialized spin-probe and detailed \textit{ab-initio} analysis to be uncovered.

\section{Conclusions}

In summary, a step-like transition in the spin-dependent resistivity of Co/Ni multilayers on Ni-doping of the Cu-spacers observed in the concentration interval 15-25 at.~\%~Ni is attributed to the Ni impurities transitioning between atomically nonmagnetic and magnetic states. This conclusion confirms the earlier predictions \cite{Ahern1958,Robbins1969} and is fully supported by our first-principle numerical calculations \cite{Kravets2012, Kravets2012a}. Importantly, the effect is observed only for the gradient-spacer configuration, where the dilute magnetic alloy is spaced from the strongly ferromagnetic Co by a thin pure Cu layer to avoid the Co-Ni proximity exchange. Our interpretation of the step-down in the GMR as due to a magnetic phase transition of the atomic Ni in Cu seen only for the proximity-free gradient-spacers differs in principle from the results on a similar system with uniformly Ni-doped Cu spacers, which were explained in terms of an impurity-modified Fermi-surface topology \cite{Parkin1993, Okuno1993, Bobo1993}. The demonstrated approach is rather general and can be used for investigating a variety of other weakly magnetic materials, such as impurity-induced magnetism in Pt-, Pd-, Cr-based alloys, antiferromagnets near N\'eel transitions, to name a few. It potentially can be employed to probe and tune spin relaxation in nanodevices based on spin-torques \cite{Ralph2008, Locatelli2014}, spin-pumping \cite{Tserkovnyak2005, Brataas2012}, and spin-thermionics \cite{Bauer2012, Kadigrobov2010, Kadigrobov2012}.

\ack{Support from the Swedish Research Council (VR Grant No.~2014-4548), the Swedish Institute Visby Programme, and the Swedish Stiftelse Olle Engkvist Byggm\"astare are gratefully acknowledged. The work is partially supported by the National Academy of Sciences of Ukraine, proj. No. 0114U000092.}

\section*{References}
\bibliographystyle{iopart-num}
\bibliography{References}

\end{document}